\documentclass[aps,prl,floatfix,twocolumn,showpacs,nofitinbib,hyperref=pdftex,citeautoscript]{revtex4-1}
\usepackage{epsfig}
\usepackage{amsmath}
\usepackage{amssymb}
\usepackage{amsfonts}
\usepackage{ dsfont }
\usepackage{ comment,color }
\usepackage{lipsum}
\usepackage{hyperref}

\hypersetup{colorlinks=true,linkcolor=blue,anchorcolor=blue,citecolor=blue,filecolor=blue,urlcolor=blue,bookmarksnumbered=true,pdfview=FitB}

\newcommand{\bk}{{{\bf{k}}}}

\newcommand{\bq}{{\bf{q}}}

\newcommand{\beqa}{\begin{eqnarray}}
\newcommand{\eeqa}{\end{eqnarray}}

\newcommand{\ua}{\uparrow}
\newcommand{\da}{\downarrow}

\begin{document}

\hsize\textwidth\columnwidth\hsize\csname@twocolumnfalse\endcsname

\title{Parity-controlled $2\pi$ Josephson effect mediated by Majorana Kramers pairs} 
\author{Constantin Schrade and Liang Fu}
\affiliation{Department of Physics, Massachusetts Institute of Technology, 77 Massachusetts Ave., Cambridge, MA 02139}

\date{\today}

\vskip1.5truecm
\begin{abstract}
We study a time-reversal-invariant topological superconductor island hosting spatially separated Majorana Kramers pairs, with weak tunnel couplings to two $s$-wave superconducting leads. When the topological superconductor island is in the Coulomb blockade regime, we predict that a Josephson current flows between the two leads due to a non-local transfer of Cooper pairs mediated by the Majorana Kramers pairs. Interestingly, we find that the sign of the Josephson current is controlled by the joint parity of all four Majorana bound states on the island. Consequently, this parity-controlled Josephson effect can be used for qubit read-out in Majorana-based quantum computing.
\end{abstract}


\maketitle

The past years have shown rapid progress towards the realization of topological superconductors (TSCs) hosting spatially separated Majorana bound states (MBSs) \cite{bib:Alicea2012,bib:Beenakker2013,bib:Lutchyn2017}, which may be useful in building a robust quantum computer. Promising platforms for TSCs to date include hybrid superconductor (SC) - semiconductor nanowire devices under magnetic fields \cite{bib:Lutchyn2010,bib:Oreg2010,bib:Mourik2012,bib:Das2012,bib:Rokhinson2012,bib:Deng2013}, chains of magnetic atoms on top of a SC substrate \cite{bib:Klinovaja2013,bib:Braunecker2013,bib:Pientka2013,bib:Nadj-Perge2013,bib:Nadj-Perge2014,bib:Ruby2015,bib:Pawlak2016} as well as vortices in SC-topological insulator heterostructures \cite{bib:Fu2008,bib:Xu2015,bib:Sun2016}.
While all of these setups are designed to search for unpaired MBSs, it was predicted that topological superconductivity also exists in time-reversal-invariant (TRI) systems and gives rise to Kramers doublets of MBSs or Majorana Kramers pairs (MKPs) \cite{bib:Schnyder2008}. In particular, a one-dimensional TRI TSC wire hosts spatially separated MKPs at its two ends. Despite consisting of two MBSs, an isolated MKP is a robust zero-energy degree of freedom protected by time reversal symmetry.

\begin{figure}[!t] \centering
\includegraphics[width=1\linewidth] {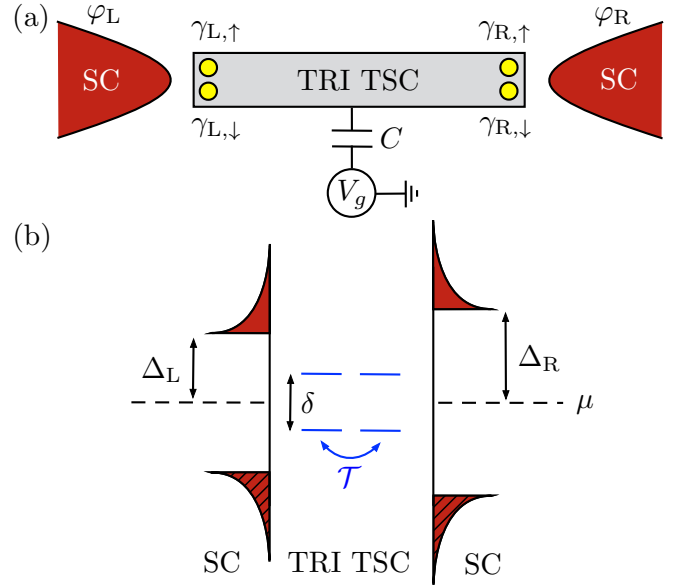}
\caption{(Color online)
(a) A superconducting weak link of two $s$-wave SC leads $\ell=\text{L,R}$ (red) with SC phases $\varphi_{\ell}$ which are coupled to a mesoscopic TRI TSC island (gray). The TRI TSC island hosts a MKP $\gamma_{\ell,s}$ with $s=\ua,\da$ (yellow) at each boundary. Moreover, the island is grounded by a capacitor with capacitance $C$ and thereby attains a finite charging energy which is tunable via an external gate voltage $V_{g}$.
(b) Schematic energy spectrum of the island and the two SC leads with superconducting gaps $\Delta_{\ell}$ close to a resonance. The low-energy charge states of the island (blue) are related by time-reversal symmetry $\mathcal{T}$ and are split by an amount $\delta$ with $|\delta|\ll\Delta_{\ell}$ due to a finite detuning away from resonance. The superconducting gap of the island is assumed to be the largest energy scale.
}\label{fig:1}
\end{figure}

Candidate systems for realizing such TRI topological superconductors comprise nanowires contacted to unconventional SCs \cite{bib:Wong2012,bib:Nagaosa2013,bib:Zhang2013,bib:Dumitrescu2014}, Josephson $\pi$-junctions in proximitized nanowires and topological insulators \cite{bib:Keselman2013,bib:Haim2014,bib:Schrade2015,bib:Borla2017} as well as setups of two nanowires or two topological insulator systems coupled via a conventional $s$-wave SC \cite{bib:Klinovaja2014,bib:Gaidamauskas2014,bib:Schrade2017,bib:Klinovaja20142}. Additionally, it was pointed out recently that TSCs could appear in systems with an emergent time-reversal symmetry \cite{bib:Huang2017,bib:Reeg2017,bib:Hu2017,bib:Maisberger2017}.
While various schemes were put forward to detect the MKPs in such systems \cite{bib:Chung2013,bib:Li2016,bib:Mellars2016,bib:Kim2016,bib:Camjayi2017,bib:Bao2017},
novel properties of MKPs and TRI TSCs remain to be explored. 

In this work, we study the Josephson effect in a mesoscopic TRI TSC island tunnel coupled to two $s$-wave superconducting leads via two spatially separated MKPs, see Fig.~\ref{fig:1}(a).
When the island is in the Coulomb blockade regime, we show that a finite Josephson current flows due to higher order co-tunnelling processes in which Cooper pairs in the SC leads tunnel in and out of the spatially separated MKPs localized at opposite ends of the island. We find that the sign of the resulting Josephson current is controlled by the joint parity of the two MKPs. For the case of odd joint parity, the two SC leads form a Josephson $\pi$-junction, whereas for even joint parity the two SC leads form a Josephson $0$-junction. Besides being a robust and easily accessible property of MKPs, we hope that the sign reversal of the Josephson current will prove useful for qubit read-out in Majorana-based quantum computing \cite{bib:Bravyi2010,bib:Vijay2015,bib:Vijay2016,bib:Landau2016,bib:Plugge2016,bib:Hoffman2016,bib:Vijay2016_2,bib:Plugge2017,bib:Karzig2016}.

{\it Model.}
We consider a TRI TSC island of mesoscopic size that is connected to the ground by a capacitor and weakly coupled to two $s$-wave SC leads, see Fig.~\ref{fig:1}(a). The two SC leads, labeled by $\ell=\text{L,R}$, are described by the BCS (Bardeen-Cooper-Schrieffer) Hamiltonian,
\begin{equation}
\label{Eq1}
H_{0}=\sum_{\ell=\text{L,R}}\sum_{\bk} \Psi_{\ell,\bk}^\dagger \left(
\xi_{\bk}\eta_{z}+\Delta_{\ell}\eta_{x}e^{i\varphi_{\ell}\eta_{z}}
\right)\Psi_{\ell,\bk}.
\end{equation}
Here, $\Psi_{\ell,\bk}=(c_{\ell,\bk\ua},c^{\dag}_{\ell,-\bk\da})^{T}$ is a Nambu spinor with
$c_{\ell,\bk s}$ the electron annihilation operator, where $\bk$ denotes single-particle states with energy $\xi_{\bk}$ in the absence of superconductivity and $s=\ua,\da$ denotes electron's spin, or more generally, Kramers degeneracy in the presence of spin-orbit coupling. By definition, $s$-wave pairing occurs between Kramers pairs $(\bk,s)$ and $(-\bk, -s)$, resulting in the superconducting gap $\Delta_\ell$. The Pauli matrices acting in Nambu space are denoted by $\eta_{x,y,z}$.
For simplicity, we assume the magnitudes of the superconducting gaps are identical, $\Delta\equiv\Delta_{\text{L}}=\Delta_{\text{R}}$. 

The TRI TSC island hosts a MKP $\gamma_{\ell,s}$ at each boundary. The two members of a MKP are related by time-reversal symmetry,
\begin{equation}
\mathcal{T}\gamma_{\ell,\ua}\mathcal{T}^{-1}=\gamma_{\ell,\da}, \;  \mathcal{T}\gamma_{\ell,\da}\mathcal{T}^{-1}=-\gamma_{\ell,\ua}.
\end{equation}
We assume that the length of the island is much longer than the MBS localization lengths, so that the wavefunction overlap of MKP on opposite boundaries is negligible. Since MBSs are zero-energy degrees of freedom that can host unpaired electrons without energy cost, the TRI TSC island is able to accommodate even and odd numbers of electrons on equal ground. For a TRI TSC island of mesoscopic size,  there is also a finite charging energy given by
\begin{equation}
U_{C}(n) = (ne-Q_{0})^{2}/2C.
\end{equation}
Here, $Q_{0}$ is a gate charge that is continuously tunable via a gate voltage $V_{g}$ across a capacitor with capacitance $C$.

Finally, we introduce the tunnel coupling between the TRI TSC island and the $s$-wave SC leads. We assume temperature is sufficiently small compared to the charging energy $U\equiv e^{2}/2C$ and the superconducting gaps of both the SC leads and the island, so that no quasiparticle states are occupied with a notable probability and the Josephson current is predominantly carried by the ground state of the junction. Moreover, we assume that the SC gap in the island is sufficiently large so that virtual transitions via quasiparticle states in the island are negligible. 

Single-particle tunneling between MKPs in the island and the SC leads is then described by the Hamiltonian
\begin{align}
\label{Eq4}
H_{T}
&=
\sum_{\ell=\text{L,R}}
\sum_{\bk,s',s}
\lambda_{\ell s s'}
c^{\dag}_{\ell,\bk s'}
\gamma_{\ell,s}
e^{-i\phi/2}
+
\text{H.c.}
\end{align}
Here, the tunneling amplitudes at the junction between the island and the lead $\lambda_{\ell s s'}$ are allowed to take the most general form, \textit{i.e.}, complex and spin-dependent. Time reversal symmetry implies
$\lambda^*_{\ell s s'} = (s_y)_{s t} \lambda_{\ell t t'} (s_y)_{t' s'}$ with $s_{x,y,z}$ denoting the Pauli matrices in spin-space.
We note that one can always choose a proper spin basis transformation so that the tunneling amplitude becomes real and spin-independent, \textit{i.e.}, $\lambda_{\ell ss'} = \lambda_{\ell} \delta_{ss'}$. Without loss of generality, this choice simplifies our analysis greatly and will be adopted below \cite{bib:supplemental}.

The operator $e^{\pm i\phi/2}$ in Eq.~\eqref{Eq4} increases/decreases the total charge of the TRI TSC island by one charge unit, $[n,e^{\pm i\phi/2}]=\pm e^{\pm i\phi/2}$ while the MBS operators $\gamma_{\ell,s}$ change the electron number parity in the TRI TSC island \cite{bib:Fu2010}. Moreover, we remark that the MBSs at one end of the island do not couple to the SC lead at the opposite end, as the localization length of the MBSs is assumed to be much shorter than the length of the island. To summarize, the full Hamiltonian of our setup is given by $H=H_{0}+U_{C}(n)+H_{T}$.

{\it Josephson current near a resonance.}
In this section, we show that a Josephson current occurs  due to Cooper pair tunnelling between the TRI TSC island and the two SC leads enabled by the two MKPs. We first focus on the near-resonant case, $|\delta|\ll\Delta$ with $\delta\equiv U_{C}(n_{0})-U_{C}(n_{0}+1)$. This allows us to truncate the Hilbert space of the island retaining only the states with $n_{0}$ and $n_{0}+1$ units of charge, see Fig.~\ref{fig:1}(b). All remaining charge states are separated from this low-energy subspace by a large charging energy, $U\gg|\delta|$, and hence have negligible contribution to the Josephson current.

\begin{figure}[!t] \centering
\includegraphics[width=1\linewidth] {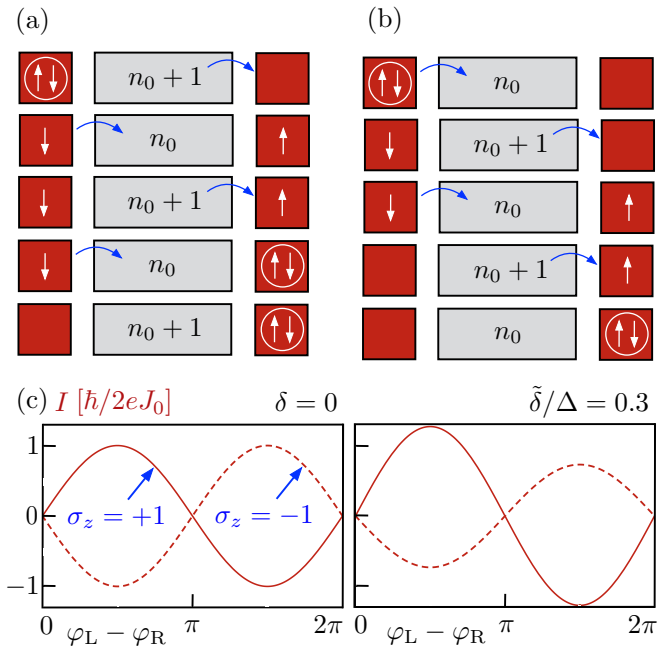}
\caption{(Color online)
(a)
Typical process for the Cooper pair transport between the SC leads via the TRI TSC island which
in this case initially carries $n_{0}+1$ units of charge.
(b)
Same as (a) but this time the island initially carries $n_{0}$ units of charge. Compared to (a), the intermediate steps
of adding/removing a charge from the TRI TSC island are reversed.
(c) 
Josephson current $I$ (in units of $\hbar/2eJ_0)$ versus the SC phase difference $\varphi_{\text{L}}-\varphi_{\text{R}}$ for $\delta=0$ (left panel) and $\tilde{\delta}/\Delta=0.3$ (right panel). If the joint parity of the MBSs is even, $\sigma_{z}=+1$, the weak link forms a Josephson $0$-junction. Otherwise, it forms a Josephson $\pi$-junction. At resonance, when $\delta=0$, the magnitude of the critical current is identical for both the even and odd parity branches (left panel). This symmetry is lifted away from resonance when $\delta\neq0$ (right panel).
}\label{fig:2}
\end{figure}

Due to the superconducting gap $\Delta$ in the SC leads, single charge transfer across the TRI TSC island is suppressed at low energy. Cooper pair transport occurring separately between each SC lead and the island is also forbidden, as these processes alter the charge of the island by $2e$ and thereby leak out of the low-energy Hilbert space. Up to fourth order in the tunneling amplitudes $\lambda_{\ell}$, only two types of co-tunnelling processes give rise to coherent Josephson coupling between the two SC leads. These processes transfer charge $2e$ between the two SC leads through the TRI TSC island. In the case of the states with $n_{0}$ electrons on the island, the transfer of a $2e$-charged Cooper pair across the junction entails four steps of subsequently adding and removing electrons on the island, see Fig.~\ref{fig:2}(a). This is also the case for states with $n_{0}+1$ electrons on the island, but the intermediate steps of adding and removing charges are reversed, see Fig.~\ref{fig:2}(b).

The amplitude of these processes at and near resonance is derived in the limit of weak tunnel coupling, $\Gamma_{\ell}\equiv\pi\nu_{\ell}|\lambda_{\ell}|^{2}\ll\Delta$ with $\nu_{\ell}$ the normal-state density of states per spin of the $\ell$-SC at the Fermi energy \cite{bib:supplemental}.
The result is summarized by an effective Hamiltonian acting on the reduced Hilbert space consisting of the BCS ground states of the SC leads and the charge states $n_{0}$ and $n_{0}+1$  of the mesoscopic TRI TSC island,
\begin{equation}
H_{\text{eff}}=
\frac{\delta}{2}\tau_{z}
-
(
\gamma_{\text{R},\ua}
\gamma_{\text{L},\ua}
\gamma_{\text{R},\da}
\gamma_{\text{L},\da}
)
\left(J_{0}+\frac{J_{1}\delta}{\Delta}\tau_{z}\right)\cos(\varphi_{\text{L}}-\varphi_{\text{R}})\label{Eq6}
\end{equation}
where $\tau_{z}=\pm 1$ denotes the charge state $n_{0}$ and $n_{0}+1$ in the island, respectively. Here, the first term describes the energy splitting $\delta$ of the two charge states due to detuning the gate charge $Q_{0}$ away from the resonant point $Q_{0}/e=n_{0}+1/2$. Moreover, $J_{0}$ is the Josephson coupling at resonance, while $J_{1}$ is the lowest-order correction for a small detuning $\delta/\Delta$ away from resonance.
Their expressions are given by
\begin{equation}
\begin{split}
J_{0}
&=
\frac{16\Gamma_{\text{L}}\Gamma_{\text{R}}}{\pi^{2}\Delta}
\int^{\infty}_{1}
\frac
{
\mathrm{d}x \text{ }�� \mathrm{d}y
}{
\left[f(x)+f(y)\right]
\left[f(x)f(y)\right]^{2}
}
\\
J_{1}
&=
\frac{16\Gamma_{\text{L}}\Gamma_{\text{R}}}{\pi^{2}\Delta}
\left(\frac{3}{2}-\sqrt{2}\right),
\end{split}
\end{equation}
where $f(x)\equiv\sqrt{1+x^{2}}$.

The effective Hamiltonian given in Eq.~(\ref{Eq6}) is the first main result of our work. Notably, it directly relates the Josephson current to the joint fermion parity of the four MBSs on the island, $\gamma_{\text{R},\ua}
\gamma_{\text{L},\ua}
\gamma_{\text{R},\da}
\gamma_{\text{L},\da}$. Depending on the fermion parity being even or odd,
$\gamma_{\text{R},\ua}
\gamma_{\text{L},\ua}
\gamma_{\text{R},\da}
\gamma_{\text{L},\da}
=\pm1$,
the Josephson current between the SC leads is given by
\begin{equation}
I=
\pm
\frac{2e}{\hbar}
\left(J_{0}+\frac{J_{1}\delta}{\Delta}\tau_{z}\right)\sin(\varphi_{\text{L}}-\varphi_{\text{R}})
\label{Current1}.
\end{equation}
Importantly, Eq.~(\ref{Eq6}) also applies to the general case of a TRI TSC island hosting any number of spatially separated MKPs. When two of these MKPs couple to separate SC leads, they mediate a Josephson current given by (\ref{Current1}) where $\pm$ denotes their joint fermion parity being even or odd, respectively.

For the simplest case of a TSC island with only two MKPs, in the absence of quasiparticle poisoning, the joint parity is given by the total island charge mod 2 \cite{bib:Fu2010,bib:Xu2010},
$
\gamma_{\text{L},\ua}\gamma_{\text{R},\ua}\gamma_{\text{L},\da}
\gamma_{\text{R},\da} = (-1)^n.
$
After truncating the Hilbert space of the island to two charge states $n=n_0, n_0+1$, the joint parity becomes
\begin{equation}
\gamma_{\text{L},\ua}\gamma_{\text{R},\ua}\gamma_{\text{L},\da}
\gamma_{\text{R},\da} = (-1)^{n_0} \tau_z.
\label{constraint1}
\end{equation}
The Josephson current given in Eq.~\eqref{Current1} then simplifies to
\begin{equation}
I=
\frac{2e}{\hbar}
\left(J_{0}\sigma_z+\frac{J_{1} \tilde{\delta}}{\Delta}\right)\sin(\varphi_{\text{L}}-\varphi_{\text{R}}),
\label{Current2}
\end{equation}
where $\sigma_{z} \equiv (-1)^{n_0} \tau_z = \pm 1$ denotes the even- (odd-) parity member of the two nearly-degenerate charge states $n_{0}$ and $n_{0}+1$ in the TRI TSC island, and $\tilde{\delta} \equiv  (-1)^{n_0} \delta$ is the energy difference of the $\sigma_z=+1$ and $\sigma_z=-1$ state.
Three aspects are noteworthy:

(1) The Josephson current between the SC leads as given in Eq.~\eqref{Current1}, or equivalently Eq.~\eqref{Current2}, is mediated solely by the MKPs localized at opposite boundaries of the TRI TSC island with its magnitude being determined by the coupling strengths at the two junctions only. This is remarkable because the MKPs have essentially zero wavefunction overlap and so no \textit{direct} coupling exists between the MKPs  in Eq.~\eqref{Eq4}.  

(2) For a given phase difference $\varphi_{\text{L}}-\varphi_{\text{R}}$, the sign of the Josephson current in Eq.(\ref{Current1}) depends on the joint fermion parity of the four MBSs.
For a given fermion parity state, the Josephson current is $2\pi$ periodic with respect to the phase.
For that reason, we refer to Eq.~\eqref{Current1} as the \textit{parity-controlled $2\pi$ Josephson effect}. When the joint parity of the four MBSs is even, $\gamma_{\text{L},\ua}\gamma_{\text{R},\ua}\gamma_{\text{L},\da}
\gamma_{\text{R},\da}= +1$, the critical current is positive, $I_{c}>0$, and the superconducting weak link forms a Josephson $0$-junction. In contrast, when the joint parity of the four MBSs is odd,  $\gamma_{\text{L},\ua}\gamma_{\text{R},\ua}\gamma_{\text{L},\da}
\gamma_{\text{R},\da}= -1$, the sign of the critical current is negative, $I_{c}<0$, and the weak link forms a Josephson $\pi$-junction, see also Fig.~\ref{fig:2}(c).

The parity-controlled Josephson effect found here has two immediate applications: First, for sufficiently long parity lifetimes, the sign of the critical current permits a direct measurement of the joint parity of \textit{four MBSs} in the island, an essential element for Majorana-based quantum computing \cite{bib:Bravyi2010,bib:Vijay2015,bib:Vijay2016,bib:Landau2016,bib:Plugge2016,bib:Hoffman2016,bib:Vijay2016_2,bib:Plugge2017,bib:Karzig2016}. In contrast, for time-reversal-breaking TSCs with unpaired MBSs, the sign of the Josephson current in the $4\pi$-periodic Josephson effect only permits measuring the parity of \textit{two MBSs} in the weak limit \cite{bib:Kitaev2001,bib:Laroche2017}. Second, the switching times between positive and negative critical currents through the island provide a way of measuring the rate of quasiparticle poisoning, which is the major source of decoherence for Majorana-based quantum bits.

(3) Eq.~\eqref{Current2} shows that on resonance ($\delta=0$), $J_{1}=0$, \textit{i.e.}, the magnitude of critical current in even and odd parity branches is identical. Away from resonance ($\delta\neq0$), $J_{1}\neq0$. Hence this symmetry is lifted and the critical current mediated by the TSC island in even or odd configurations differs in magnitude. When the even parity state is higher (lower) in energy $\tilde{\delta}>0$ ($<0$), the corresponding critical current is larger (smaller) in magnitude, see Eq.~\eqref{Current2} and Fig.~\ref{fig:2}(c). 

{\it Josephson current near a Coulomb valley.}
So far, we have restricted our discussion to the case when the gate charge is tuned close to resonance, $|\delta|\ll\Delta$. In this section, we show that the proposed parity-controlled Josephson effect is more general and also arises near a Coulomb valley when $Q_{0}/e$ is close to an integer value, $2N+1$ or $2N$, so that the ground states of the island consist of either an odd number of electrons, $n_{0}=2N+1$, or an even number of electrons, $n_{0}=2N$.

Under this condition, Cooper pair transport occurs microscopically via virtually excited states of order $U$ on the island.  Up to fourth order in the tunnelling amplitudes $\lambda_{\ell}$, three types of co-tunnelling processes contribute to the Josephson coupling: The first type of process involves subsequently adding and removing a unit of charge on the island, similar to the processes discussed for the close-to-resonance case. For the second type of process, the first two intermediate steps involve adding/removing a charge on the island, while in the final two intermediate steps this order of adding/removing a charge is reversed. In the third type of process, a Cooper pair from one lead is added/removed on the island in the first two intermediate steps, which alters the island charge by $2e$. Subsequently, the Cooper pair is again removed/added from/to the other lead in the final two intermediate steps so that the island returns to its ground state. Importantly, the processes of the second and third type involve intermediate charge states $n_0 - 1, n_0\pm 2$, which are energetically unfavourable in the close-to-resonance case, but in the Coulomb valley case, should be included.

The amplitudes of the processes described above can be calculated in the limit of weak tunnel couplings, $\Gamma_{\ell}\ll\Delta, U$, using fourth-order perturbation theory. The result is summarized in the form of an effective Hamiltonian acting on the BCS ground states of the SC leads and the charge ground states on the island,
\begin{equation}
\begin{split}
H'_{\text{eff}}=
-(
\gamma_{\text{R},\ua}
\gamma_{\text{L},\ua}
\gamma_{\text{R},\da}
\gamma_{\text{L},\da}
)J'\cos(\varphi_{\text{L}}-\varphi_{\text{R}}).
\label{Eq9}
\end{split}
\end{equation}
Here, we have introduced the coupling constant $J'=J'_{0}+J'_{1}+J'_{2}$ with,
\begin{align}
J'_{0}
&=
\frac{32\Gamma_{\text{L}}\Gamma_{\text{R}}}{\pi^{2}\Delta}
\int^{\infty}_{1}
\frac
{
\mathrm{d}x \text{ }�� \mathrm{d}y
}{
f(x)f(y)
\left[f(x)+f(y)\right]g(x)g(y)}\nonumber
\\
J'_{1}
&=
\frac{32\Gamma_{\text{L}}\Gamma_{\text{R}}}{\pi^{2}\Delta}
\int^{\infty}_{1}
\frac
{
\mathrm{d}x \text{ }�� \mathrm{d}y
}{
f(x)f(y)
\left[f(x)+f(y)\right]g(x)^{2}}
\\
J'_{2}
&=
\frac{8\Gamma_{\text{L}}\Gamma_{\text{R}}}{\pi^{2} U}
\left[
\int^{\infty}_{1}
\frac
{
\mathrm{d}x \text{ }��
}{
f(x)g(x)}
\right]^{2},
\nonumber
\end{align}
and $g(x)\equiv\sqrt{1+x^{2}}+U/\Delta$. The effective Hamiltonian given in Eq.~(\ref{Eq9}) is the second main result of our work.

Crucially, we observe that the direct coupling of the effective Hamiltonian to the joint parity $\gamma_{\text{R},\ua}
\gamma_{\text{L},\ua}
\gamma_{\text{R},\da}
\gamma_{\text{L},\da}$ is preserved near a Coulomb valley.
For the simplest case when the joint parity is fixed by the total island charge mod 2,
\begin{equation}
\gamma_{\text{L},\ua}\gamma_{\text{R},\ua}\gamma_{\text{L},\da}
\gamma_{\text{R},\da} = (-1)^{n_0}.
\label{Constraint2}
\end{equation}
the resulting Josephson current is given by,
 \begin{equation}
I'=(-1)^{n_{0}}(2e/\hbar)J'\sin(\varphi_{\text{L}}-\varphi_{\text{R}}).
\label{Eq11}
\end{equation}
We want to emphasize three features of this result:

(1) Unlike in the close-to-resonance case, the Josephson current consists of only a single branch for either an even parity ground state, $n_{0}=2N$, or an odd parity ground state, $n_{0}=2N+1$. However, the sign of the critical current $I'_{c}\equiv(-1)^{n_{0}}(2e/\hbar)J'$ remains to be a direct measure of the joint parity $\gamma_{\text{R},\ua}
\gamma_{\text{L},\ua}
\gamma_{\text{R},\da}
\gamma_{\text{L},\da}$ through the gauge constraint given in Eq.~\eqref{Constraint2}.

(2) In comparison to the close-to-resonance case, the sign of the supercurrent is expected to be more stable against quasiparticle poisoning events due to the large charging energy.

(3) At the Coulomb valleys, the magnitude of the critical current is identical for both even and odd configurations. This behavior is in contrast with weak links of two SC leads coupled via a quantum dot, where odd and even charge states of the quantum dot create Josephson $0$- and $\pi$-junctions, respectively, but with critical current generally of different magnitude \cite{bib:Kouwenhoven2006}.

Before closing, we point out that under rather general conditions no Josephson current is observed when the TRI TSC island
is replaced by a time-reversal-breaking TSC island (in symmetry class $D$) \cite{bib:Zazunov2012}. This is because after a proper spin basis transformation,
a non-degenerate MBS in the TSC island couple only to a single spin species \cite{bib:Law2009} and not to both spin species as MKPs do in the case for a TRI TSC island. 

{\it Conclusions.}
We have shown that in a weak link of two $s$-wave SCs coupled via a TRI TSC island, a Josephson current can flow
due to Cooper pairs tunneling in and out of spatially separated MKPs.
We have demonstrated that the sign of the resulting Josephson current is fixed by the joint parity of
the four MBSs on the island. As a consequence, this  \textit{parity-controlled Josephson effect} can be used as a read-out mechanism for the joint parity, which is a key requirement in Majorana-based quantum computing \cite{bib:Bravyi2010,bib:Vijay2015,bib:Vijay2016,bib:Landau2016,bib:Plugge2016,bib:Hoffman2016,bib:Vijay2016_2,bib:Plugge2017,bib:Karzig2016}.

{\it Acknowledgments.}
We would like to thank Patrick A. Lee for helpful discussions. C.S. was supported by the Swiss SNF under Project 174980. L.F. and C.S. were supported by DOE Office of Basic Energy Sciences, Division of Materials Sciences and Engineering under Award $\text{DE-SC0010526}$.

\begin{widetext}

\newpage

\onecolumngrid

\bigskip

\begin{center}
\large{\bf Supplemental Material to `Parity-controlled $2\pi$ Josephson effect mediated by Majorana Kramers pairs' \\}
\end{center}
\begin{center}
Constantin Schrade and Liang Fu
\\
{\it Department of Physics, Massachusetts Institute of Technology, 77 Massachusetts Ave., Cambridge, MA 02139}
\end{center}

In the Supplemental Material, we provide more details on the derivation of tunneling Hamiltonian coupling the SC leads to the TRI TSC island as well as on the derivation of the effective Hamiltonians describing the parity-controlled Josephson effect.

\section{Tunneling Hamiltonian}
In this first section of the Supplemental Material, we derive the tunneling Hamiltonian between the SC leads and the TRI TSC island. As a starting point, we note that the constraint $\lambda^*_{\ell s s'} = (s_y)_{s t} \lambda_{\ell t t'} (s_y)_{t' s'}$ on the tunneling amplitudes due to time-reversal symmetry is equivalent to the requirement that $\lambda_{\ell \ua\ua}=\lambda^*_{\ell \da\da}$ and $\lambda_{\ell \ua\da}=-\lambda^*_{\ell \da\ua}$. Defining new tunneling amplitudes $t_{\ell}\equiv\lambda_{\ell \ua\ua}$ and $\tilde{t}_{\ell}\equiv\lambda_{\ell \ua\da}$ allows us to rewrite the most general time-reversal symmetric coupling between the fermions in the SC leads and the MKPs in the TRI TSC island as,
\begin{align}
\label{TunnelingH1}
H_{T}
&=
\sum_{\ell=\text{L,R}}
\sum_{\bk}
\Big[
\Big(
t_{\ell}
c^{\dag}_{\ell,\bk \ua}
\gamma_{\ell,\ua}
+
t^{*}_{\ell}
c^{\dag}_{\ell,\bk \da}
\gamma_{\ell,\da}
\Big)
e^{-i\phi/2}
+
\Big(
\tilde t_{\ell}
c^{\dag}_{\ell,\bk \ua}
\gamma_{\ell,\da}
-
\tilde t^{*}_{\ell}
c^{\dag}_{\ell,\bk\da}
\gamma_{\ell,\ua}
\Big)e^{-i\phi/2}+\text{H.c.}\Big].
\end{align}
We remark that for simplicity, we have assumed a point-like tunneling contact so that the complex tunneling amplitudes $t_{\ell}, \tilde{t}_{\ell}$ are momentum-independent.
Next, we perform a unitary rotation in the space of the fermions in the SC grains,
\begin{equation}
\label{rotation}
\begin{pmatrix}
d_{\ell,\bk \ua} \\
d_{\ell,-\bk \da}
\end{pmatrix}
\equiv
\frac{1}{
\sqrt{
|t_{\ell}|^{2}+|\tilde{t}_{\ell}|^{2}
}
}
\begin{pmatrix}
t^{*}_{\ell} & -\tilde{t}_{\ell} \\
\tilde{t}^{*}_{\ell} & t_{\ell}
\end{pmatrix}
\cdot
\begin{pmatrix}
c_{\ell,\bk \ua} \\
c_{\ell,-\bk \da}
\end{pmatrix}.
\end{equation}
We remark that in the newly introduced fermionic operators $d_{\ell,\bk s}$ the label $s=\ua,\da$ refers to a Kramers index
defined via $\mathcal{T}d_{\ell,\bk \ua}\mathcal{T}^{-1}=d_{\ell,-\bk \da}$ and $\mathcal{T}d_{\ell,-\bk \da}\mathcal{T}^{-1}=-d_{\ell,\bk \ua}$. Using the unitary transformation in Eq.~\eqref{rotation}, the tunneling Hamiltonian in Eq.~\eqref{TunnelingH1} transforms to
\begin{align}
\label{TunnelingH2}
H_{T}
&=
\sum_{\ell=\text{L,R}}
\sum_{\bk,s}
\sqrt{
|t_{\ell}|^{2}+|\tilde{t}_{\ell}|^{2}
}\hspace{2pt}
d^{\dag}_{\ell,\bk s}
\gamma_{\ell,s}
e^{-i\phi/2}
+\text{H.c.}
\end{align}
With the appropriate relabelling, $\sqrt{|t_{\ell}|^{2}+|\tilde{t}_{\ell}|^{2}}\rightarrow\lambda_{\ell}$ and $d_{\ell,\bk s}\rightarrow c_{\ell,\bk s}$, we have thus reproduced the tunneling Hamiltonian of Eq.~(4) in the main text with the additional condition that $\lambda_{\ell s s'}=\lambda_{\ell}\delta_{ss'}$.

\section{Effective Hamiltonian close to resonance}
In this second section of the Supplemental Material, we provide the derivation of the effective Hamiltonian for the parity-controlled Josephson effect for the case when the gate charge $Q_{0}/e$ is tuned close to resonance so that the states with charges $n_{0}$ and $n_{0}+1$ are almost degenerate, $|\delta|\ll\Delta$.
Up to fourth order in the tunnel couplings, the general form of the effective Hamiltonian is then given by
\begin{align}
H_{\text{eff}}&=
P_{n_{0}} U_{C}(n) P_{n_{0}}
- P_{n_{0}} H_{T} \left(\left[H_{0}+U_{C}(n)-U_{C}(n_{0})\right]^{-1}\left[1-P_{n_{0}}\right]H_{T}\right)^{3}P_{n_{0}} \\
&\quad\
P_{n_{0}+1} U_{C}(n) P_{n_{0}+1}
- P_{n_{0}+1} H_{T} \left(\left[H_{0}+U_{C}(n)-U_{C}(n_{0}+1)\right]^{-1}\left[1-P_{n_{0}+1}\right]H_{T}\right)^{3}P_{n_{0}+1}.
\nonumber
\label{EffH_Close_to_resonance}
\end{align}
Here, we have omitted the second order contribution as it only leads to a constant shift in energy and consequently does not contribute to the Josephson current.  Moreover, $P_{n}=\Pi_{n}\Pi_{\text{BCS}}$ with $\Pi_{n}$ a projector on the states with $n$ electrons on the mesoscopic TRI TSC and $\Pi_{\text{BCS}}$ a projector on the BCS ground states of the SC leads.
As a next step, we evaluate the effective Hamiltonian by collecting the various sequences of
intermediate states. To explain this procedure, we first consider an example for a sequence of intermediate states
that starts in a state with $n_{0}$ electrons on the TRI TS island,
\begin{equation}
\begin{split}
&\quad\
P_{n_{0}}
(
\lambda_{\text{R}}
c^{\dag}_{\text{R},\bq \ua}
\gamma_{\text{R},\ua}
e^{-i\phi/2}
)
(
\lambda_{\text{L}}
\gamma_{\text{L},\ua}
c_{\text{L},\bk \ua}
e^{i\phi/2}
)
(
\lambda_{\text{R}}
c^{\dag}_{\text{R},-\bq \da}
\gamma_{\text{R},\da}
e^{-i\phi/2}
)
(
\lambda_{\text{L}}
\gamma_{\text{L},\da}
c_{\text{L},-\bk \da}
e^{i\phi/2}
)
P_{n_{0}}
\\
&=
\lambda_{\text{R}}^{2}
\lambda_{\text{L}}^{2} \
P_{n_{0}}
(
c^{\dag}_{\text{R},\bq \ua}
\gamma_{\text{R},\ua}
\gamma_{\text{L},\ua}
c_{\text{L},\bk \ua}
c^{\dag}_{\text{R},-\bq \da}
\gamma_{\text{R},\da}
\gamma_{\text{L},\da}
c_{\text{L},-\bk \da}
)
P_{n_{0}}
\\
&=
\lambda_{\text{R}}^{2}
\lambda_{\text{L}}^{2}
\
\Pi_{n_0}
(
\gamma_{\text{R},\ua}
\gamma_{\text{L},\ua}
\gamma_{\text{R},\da}
\gamma_{\text{L},\da}
)
\Pi_{n_0}
\
\Pi_{\text{BCS}}
(
c^{\dag}_{\text{R},\bq \ua}
c_{\text{L},\bk \ua}
c^{\dag}_{\text{R},-\bq \da}
c_{\text{L},-\bk \da}
)
\Pi_{\text{BCS}}
\\
&=
-
e^{i(\varphi_{\text{L}}-\varphi_{\text{R}})}
\lambda_{\text{R}}^{2}
\lambda_{\text{L}}^{2}
v_{\bq}
u_{\bk}
u_{\bq}
v_{\bk}
\
\Pi_{n_0}(
\gamma_{\text{R},\ua}
\gamma_{\text{L},\ua}
\gamma_{\text{R},\da}
\gamma_{\text{L},\da}
)
\Pi_{n_0}
\
\Pi_{\text{BCS}}
(
\gamma_{\text{R},-\bq \da}
\gamma_{\text{L},\bk \ua}
\gamma^{\dag}_{\text{R},-\bq \da}
\gamma^{\dag}_{\text{L},\bk \ua}
)
\Pi_{\text{BCS}}
\\
&=
\Pi_{n_0}
(
\gamma_{\text{R},\ua}
\gamma_{\text{L},\ua}
\gamma_{\text{R},\da}
\gamma_{\text{L},\da}
)
\Pi_{n_0}
e^{i(\varphi_{\text{L}}-\varphi_{\text{R}})}
\lambda_{\text{R}}^{2}
\lambda_{\text{L}}^{2}
v_{\bq}
u_{\bk}
u_{\bq}
v_{\bk} \
\Pi_{\text{BCS}}
\label{Example1}
\end{split}
\end{equation}
We remark that in the fourth
equality we have expressed the electron operators of the SC grains in terms of Bogoliubov quasiparticles,
$c_{\ell,\bk \ua}=e^{i\varphi_{\ell}/2}(u_{\bk}\gamma_{\ell,\bk \ua}+v_{\bk}\gamma^{\dag}_{\ell,-\bk \da})$ and $c_{\ell,-\bk \da}=e^{i\varphi_{\ell}/2}(u_{\bk}\gamma_{\ell,-\bk \da}-v_{\bk}\gamma^{\dag}_{\ell,\bk \ua})$. Moreover, we note
that there are three additional sequences that yield the same result as the sequence given in Eq.~\eqref{Example1},
\begin{equation}
\begin{split}
&
P_{n_{0}}
(
\lambda_{\text{R}}
c^{\dag}_{\text{R},-\bq \da}
\gamma_{\text{R},\da}
e^{-i\phi/2}
)
(
\lambda_{\text{L}}
\gamma_{\text{L},\ua}
c_{\text{L},\bk \ua}
e^{i\phi/2}
)
(
\lambda_{\text{R}}
c^{\dag}_{\text{R},\bq \ua}
\gamma_{\text{R},\ua}
e^{-i\phi/2}
)
(
\lambda_{\text{L}}
\gamma_{\text{L},\da}
c_{\text{L},-\bk \da}
e^{i\phi/2}
)
P_{n_{0}},
\\
&
P_{n_{0}}
(
\lambda_{\text{R}}
c^{\dag}_{\text{R},\bq \ua}
\gamma_{\text{R},\ua}
e^{-i\phi/2}
)
(
\lambda_{\text{L}}
\gamma_{\text{L},\da}
c_{\text{L},-\bk \da}
e^{i\phi/2}
)
(
\lambda_{\text{R}}
c^{\dag}_{\text{R},-\bq \da}
\gamma_{\text{R},\da}
e^{-i\phi/2}
)
(
\lambda_{\text{L}}
\gamma_{\text{L},\ua}
c_{\text{L},\bk \ua}
e^{i\phi/2}
)
P_{n_{0}},
\\
&
P_{n_{0}}
(
\lambda_{\text{R}}
c^{\dag}_{\text{R},-\bq \da}
\gamma_{\text{R},\da}
e^{-i\phi/2}
)
(
\lambda_{\text{L}}
\gamma_{\text{L},\da}
c_{\text{L},-\bk \da}
e^{i\phi/2}
)
(
\lambda_{\text{R}}
c^{\dag}_{\text{R},\bq \ua}
\gamma_{\text{R},\ua}
e^{-i\phi/2}
)
(
\lambda_{\text{L}}
\gamma_{\text{L},\ua}
c_{\text{L},\bk \ua}
e^{i\phi/2}
)P_{n_{0}}.
\end{split}
\end{equation}
If we combine these findings with the results from the corresponding hermitian-conjugated sequences, multiply by the respective energy denominators and perform the summation over all momenta, we find that
\begin{equation}
P_{n_{0}}H_{\text{eff}}P_{n_{0}}
=
\frac{\delta}{2}P_{n_{0}}- P_{n_{0}}8 (
\gamma_{\text{R},\ua}
\gamma_{\text{L},\ua}
\gamma_{\text{R},\da}
\gamma_{\text{L},\da}
)
\lambda_{\text{R}}^{2}
\lambda_{\text{L}}^{2}
\cos(\varphi_{\text{L}}-\varphi_{\text{R}})
\sum_{\bk,\bq}
\frac{
v_{\bq}
u_{\bk}
u_{\bq}
v_{\bk}
}
{
(E_{\bq}-\delta)
(E_{\bk}+E_{\bq})
(E_{\bk}-\delta)
}P_{n_{0}}.
\label{Example1_Collected}
\end{equation}
As a next step, we simplify this expression by rewriting the summation over the momenta in terms of an integral
over the density of states. This yields
\begin{equation}
\begin{split}
&
P_{n_{0}}H_{\text{eff}}P_{n_{0}}
=
\frac{\delta}{2}P_{n_{0}}-P_{n_{0}}
(
\gamma_{\text{R},\ua}
\gamma_{\text{L},\ua}
\gamma_{\text{R},\da}
\gamma_{\text{L},\da}
)
\frac{16\Gamma_{\text{L}}\Gamma_{\text{R}}\cos(\varphi_{\text{L}}-\varphi_{\text{R}})}{\pi^{2}\Delta}
\int^{\infty}_{1}
\frac
{
\mathrm{d}x \text{ }�� \mathrm{d}y
}{
f(x)f(y)
\left[f(x)+f(y)\right]h_{-}(x)h_{-}(y)
}P_{n_{0}},
\label{Example1_Collected2}
\end{split}
\end{equation}
where we have defined $f(x)=\sqrt{1+x^{2}}$, $h_{\pm}(x)=\sqrt{1+x^{2}}\pm\delta/\Delta$. As we have assumed that $|\delta|\ll\Delta$, we expand the integrand on the right-hand side to first order in $\delta/\Delta$. This gives
\begin{equation}
P_{n_{0}}H_{\text{eff}}P_{n_{0}}
=
\frac{\delta}{2}P_{n_{0}}-P_{n_{0}}
(
\gamma_{\text{R},\ua}
\gamma_{\text{L},\ua}
\gamma_{\text{R},\da}
\gamma_{\text{L},\da}
)
\left(J_{0} + \frac{J_{1}\delta}{\Delta}\right)
\cos(\varphi_{\text{L}}-\varphi_{\text{R}})
P_{n_{0}},
\label{EffH_Final1}
\end{equation}
where we have introduced the coupling constants $J_{0}$ and $J_{1}$ as given in Eq.~(6) of the main text.

As a next step, we change our focus to the sequences of intermediate states that start in a state with $n_{0}+1$ electrons on the TRI TS island. An example of such a sequence is given by,
\begin{align}
&\quad\
P_{n_{0}+1}
(
\lambda_{\text{R}}
\gamma_{\text{R},\ua}
c_{\text{R},\bk \ua}
e^{i\phi/2}
)
(
\lambda_{\text{L}}
c^{\dag}_{\text{L},\bq \ua}
\gamma_{\text{L},\ua}
e^{-i\phi/2}
)
(
\lambda_{\text{R}}
\gamma_{\text{R},\da}
c_{\text{R},-\bk \da}
e^{i\phi/2}
)
(
\lambda_{\text{L}}
c^{\dag}_{\text{L},-\bq \da}
\gamma_{\text{L},\da}
e^{-i\phi/2}
)
P_{n_{0}+1}
\nonumber
\\
&=
\lambda_{\text{R}}^{2}
\lambda_{\text{L}}^{2}\
P_{n_{0}+1}
(
\gamma_{\text{R},\ua}
c_{\text{R},\bk \ua}
c^{\dag}_{\text{L},\bq \ua}
\gamma_{\text{L},\ua}
\gamma_{\text{R},\da}
c_{\text{R},-\bk \da}
c^{\dag}_{\text{L},-\bq \da}
\gamma_{\text{L},\da}
)
P_{n_{0}+1}
\nonumber
\\
&=
\lambda_{\text{R}}^{2}
\lambda_{\text{L}}^{2}\
\Pi_{n_{0}+1}
(
\gamma_{\text{R},\ua}
\gamma_{\text{L},\ua}
\gamma_{\text{R},\da}
\gamma_{\text{L},\da}
)
\Pi_{n_{0}+1}
\
\Pi_{\text{BCS}}
(
c^{\dag}_{\text{L},\bq \ua}
c_{\text{R},\bk \ua}
c^{\dag}_{\text{L},-\bq \da}
c_{\text{R},-\bk \da}
)
\Pi_{\text{BCS}}
\\
&=
\Pi_{n_{0}+1}
(
\gamma_{\text{R},\ua}
\gamma_{\text{L},\ua}
\gamma_{\text{R},\da}
\gamma_{\text{L},\da}
)
\Pi_{n_{0}+1}
e^{i(\varphi_{\text{L}}-\varphi_{\text{R}})}
\lambda_{\text{R}}^{2}
\lambda_{\text{L}}^{2}
v_{\bq}
u_{\bk}
u_{\bq}
v_{\bk}\
\Pi_{\text{BCS}}.
\nonumber
\end{align}
Again, there are three additional sequences which lead to the same result. They are given by
\begin{equation}
\begin{split}
&
P_{n_{0}+1}
(
\lambda_{\text{R}}
\gamma_{\text{R},\da}
c_{\text{R},-\bk \da}
e^{i\phi/2}
)
(
\lambda_{\text{L}}
c^{\dag}_{\text{L},\bq \ua}
\gamma_{\text{L},\ua}
e^{-i\phi/2}
)
(
\lambda_{\text{R}}
\gamma_{\text{R},\ua}
c_{\text{R},\bk \ua}
e^{i\phi/2}
)
(
\lambda_{\text{L}}
c^{\dag}_{\text{L},-\bq \da}
\gamma_{\text{L},\da}
e^{-i\phi/2}
)P_{n_{0}+1},
\\
&P_{n_{0}+1}
(
\lambda_{\text{R}}
\gamma_{\text{R},\ua}
c_{\text{R},\bk \ua}
e^{i\phi/2}
)
(
\lambda_{\text{L}}
c^{\dag}_{\text{L},-\bq \da}
\gamma_{\text{L},\da}
e^{-i\phi/2}
)
(
\lambda_{\text{R}}
\gamma_{\text{R},\da}
c_{\text{R},-\bk \da}
e^{i\phi/2}
)
(
\lambda_{\text{L}}
c^{\dag}_{\text{L},\bq \ua}
\gamma_{\text{L},\ua}
e^{-i\phi/2}
)P_{n_{0}+1},
\\
&P_{n_{0}+1}
(
\lambda_{\text{R}}
\gamma_{\text{R},\da}
c_{\text{R},-\bk \da}
e^{i\phi/2}
)
(
\lambda_{\text{L}}
c^{\dag}_{\text{L},-\bq \da}
\gamma_{\text{L},\da}
e^{-i\phi/2}
)
(
\lambda_{\text{R}}
\gamma_{\text{R},\ua}
c_{\text{R},\bk \ua}
e^{i\phi/2}
)
(
\lambda_{\text{L}}
c^{\dag}_{\text{L},\bq \ua}
\gamma_{\text{L},\ua}
e^{-i\phi/2}
)P_{n_{0}+1}.
\end{split}
\end{equation}
If we once more combine these results with the ones from the corresponding hermitian-conjugated sequences, multiply by the respective energy denominators and perform the summation over all momenta, we arrive at
\begin{equation}
P_{n_{0}+1}H_{\text{eff}}P_{n_{0}+1}
=
-\frac{\delta}{2}P_{n_{0}+1}-P_{n_{0}+1}
8(
\gamma_{\text{R},\ua}
\gamma_{\text{L},\ua}
\gamma_{\text{R},\da}
\gamma_{\text{L},\da}
)
\lambda_{\text{R}}^{2}
\lambda_{\text{L}}^{2}
\cos(\varphi_{\text{L}}-\varphi_{\text{R}})
\sum_{\bk,\bq}
\frac{
v_{\bq}
u_{\bk}
u_{\bq}
v_{\bk}
}
{
(E_{\bq}+\delta)
(E_{\bk}+E_{\bq})
(E_{\bk}+\delta)
}
P_{n_{0}+1}
.
\label{Example2_Collected}
\end{equation}
We now proceed by rewriting the summation over the momenta as an integral over the density of states. The
expression above then simplifies to
\begin{align}
P_{n_{0}+1}H_{\text{eff}}P_{n_{0}+1}
&=
-\frac{\delta}{2}P_{n_{0}+1}
\\
&\quad-P_{n_{0}+1}(
\gamma_{\text{R},\ua}
\gamma_{\text{L},\ua}
\gamma_{\text{R},\da}
\gamma_{\text{L},\da}
)
\frac{16
\Gamma_{\text{L}}\Gamma_{\text{R}}\cos(\varphi_{\text{L}}-\varphi_{\text{R}})}{\pi^{2}\Delta}
\int^{\infty}_{1}
\frac
{
\mathrm{d}x \text{ }�� \mathrm{d}y
}{
f(x)f(y)
\left[f(x)+f(y)\right]h_{+}(x)h_{+}(y)
}P_{n_{0}+1}\nonumber,
\label{Example2_Collected2}
\end{align}
Since $|\delta|\ll\Delta$, we can expand the integrand on the right-hand side to first order in  $\delta/\Delta$.
This leads us to
\begin{equation}
P_{n_{0}+1}H_{\text{eff}}P_{n_{0}+1}
=
-
\frac{\delta}{2}P_{n_{0}+1}
-P_{n_{0}+1}
(
\gamma_{\text{R},\ua}
\gamma_{\text{L},\ua}
\gamma_{\text{R},\da}
\gamma_{\text{L},\da}
)
\left(J_{0} - \frac{J_{1}\delta}{\Delta}\right)
\cos(\varphi_{\text{L}}-\varphi_{\text{R}})
P_{n_{0}+1}
\label{EffH_Final2}
.
\end{equation}
Finally, we can combine the finding of Eq.~\eqref{EffH_Final1} and Eq.~\eqref{EffH_Final2} to arrive at
\begin{equation}
\begin{split}
H_{\text{eff}}=
\frac{\delta}{2}\tau_{z}
-
(
\gamma_{\text{R},\ua}
\gamma_{\text{L},\ua}
\gamma_{\text{R},\da}
\gamma_{\text{L},\da}
)
(J_{0}+\frac{J_{1}\delta}{\Delta}\tau_{z})\cos(\varphi_{\text{L}}-\varphi_{\text{R}}),
\end{split}
\end{equation}
where $\tau_{z}=\pm1$ for states with $n_{0}$ and $n_{0}+1$ electrons on the TRI TSC island, respectively.
This concludes the derivation of the effective Hamiltonian close to a resonance.

\section{Effective Hamiltonian near a Coulomb valley}
In this third section of the Supplemental Material, we give the derivation of the effective Hamiltonian for the parity-controlled Josephson effect for the case when the gate charge $Q_{0}/e$ is tuned close to an integer, $2N$ or $2N+1$. In this case, the ground state will consist of either $n_{0}=2N$ or $n_{0}=2N+1$ electrons. Moreover, the cost of adding/removing a single electron from the island is approximately given by $U\equiv e^{2}/2C$, while the cost of adding/removing two electrons is approximately given by $4U$. Up to fourth order in the tunnel couplings, the general form of the effective Hamiltonian reads,
\begin{equation}
\begin{split}
H'_{\text{eff}}&=
- P_{n_{0}} H_{T} \left(\left[H_{0}+U_{C}(n)\right]^{-1}\left[1-P_{n_{0}}\right]H_{T}\right)^{3}P_{n_{0}}.
\end{split}
\end{equation}
As in the previous section, we have omitted the second order contribution since it only yields a constant energy shift
and hence does not contribute to the Josephson current. Furthermore, $P_{n_0}=\Pi_{n_0}\Pi_{\text{BCS}}$ with $\Pi_{n_0}$ a projector on the states with $n_0$ units of charge on the mesoscopic TRI TSC and $\Pi_{\text{BCS}}$ a projector on the BCS ground states of the SC leads. We now proceed by evaluating the sequences of intermediate states which make
up the effective Hamiltonian. First, we note that the sequences of intermediate states
\begin{equation}
\begin{split}
&
P_{n_{0}}
(
\lambda_{\text{R}}
c^{\dag}_{\text{R},\bq \ua}
\gamma_{\text{R},\ua}
e^{-i\phi/2}
)
(
\lambda_{\text{L}}
\gamma_{\text{L},\ua}
c_{\text{L},\bk \ua}
e^{i\phi/2}
)
(
\lambda_{\text{R}}
c^{\dag}_{\text{R},-\bq \da}
\gamma_{\text{R},\da}
e^{-i\phi/2}
)
(
\lambda_{\text{L}}
\gamma_{\text{L},\da}
c_{\text{L},-\bk \da}
e^{i\phi/2}
)
P_{n_{0}}
\\
&
P_{n_{0}}
(
\lambda_{\text{R}}
c^{\dag}_{\text{R},-\bq \da}
\gamma_{\text{R},\da}
e^{-i\phi/2}
)
(
\lambda_{\text{L}}
\gamma_{\text{L},\ua}
c_{\text{L},\bk \ua}
e^{i\phi/2}
)
(
\lambda_{\text{R}}
c^{\dag}_{\text{R},\bq \ua}
\gamma_{\text{R},\ua}
e^{-i\phi/2}
)
(
\lambda_{\text{L}}
\gamma_{\text{L},\da}
c_{\text{L},-\bk \da}
e^{i\phi/2}
)
P_{n_{0}},
\\
&
P_{n_{0}}
(
\lambda_{\text{R}}
c^{\dag}_{\text{R},\bq \ua}
\gamma_{\text{R},\ua}
e^{-i\phi/2}
)
(
\lambda_{\text{L}}
\gamma_{\text{L},\da}
c_{\text{L},-\bk \da}
e^{i\phi/2}
)
(
\lambda_{\text{R}}
c^{\dag}_{\text{R},-\bq \da}
\gamma_{\text{R},\da}
e^{-i\phi/2}
)
(
\lambda_{\text{L}}
\gamma_{\text{L},\ua}
c_{\text{L},\bk \ua}
e^{i\phi/2}
)
P_{n_{0}},
\\
&
P_{n_{0}}
(
\lambda_{\text{R}}
c^{\dag}_{\text{R},-\bq \da}
\gamma_{\text{R},\da}
e^{-i\phi/2}
)
(
\lambda_{\text{L}}
\gamma_{\text{L},\da}
c_{\text{L},-\bk \da}
e^{i\phi/2}
)
(
\lambda_{\text{R}}
c^{\dag}_{\text{R},\bq \ua}
\gamma_{\text{R},\ua}
e^{-i\phi/2}
)
(
\lambda_{\text{L}}
\gamma_{\text{L},\ua}
c_{\text{L},\bk \ua}
e^{i\phi/2}
)P_{n_{0}},
\\
\\
&
P_{n_{0}}
(
\lambda_{\text{R}}
\gamma_{\text{R},\ua}
c_{\text{R},\bk \ua}
e^{i\phi/2}
)
(
\lambda_{\text{L}}
c^{\dag}_{\text{L},\bq \ua}
\gamma_{\text{L},\ua}
e^{-i\phi/2}
)
(
\lambda_{\text{R}}
\gamma_{\text{R},\da}
c_{\text{R},-\bk \da}
e^{i\phi/2}
)
(
\lambda_{\text{L}}
c^{\dag}_{\text{L},-\bq \da}
\gamma_{\text{L},\da}
e^{-i\phi/2}
)
P_{n_{0}}
\\
&
P_{n_{0}}
(
\lambda_{\text{R}}
\gamma_{\text{R},\da}
c_{\text{R},-\bk \da}
e^{i\phi/2}
)
(
\lambda_{\text{L}}
c^{\dag}_{\text{L},\bq \ua}
\gamma_{\text{L},\ua}
e^{-i\phi/2}
)
(
\lambda_{\text{R}}
\gamma_{\text{R},\ua}
c_{\text{R},\bk \ua}
e^{i\phi/2}
)
(
\lambda_{\text{L}}
c^{\dag}_{\text{L},-\bq \da}
\gamma_{\text{L},\da}
e^{-i\phi/2}
)P_{n_{0}},
\\
&P_{n_{0}}
(
\lambda_{\text{R}}
\gamma_{\text{R},\ua}
c_{\text{R},\bk \ua}
e^{i\phi/2}
)
(
\lambda_{\text{L}}
c^{\dag}_{\text{L},-\bq \da}
\gamma_{\text{L},\da}
e^{-i\phi/2}
)
(
\lambda_{\text{R}}
\gamma_{\text{R},\da}
c_{\text{R},-\bk \da}
e^{i\phi/2}
)
(
\lambda_{\text{L}}
c^{\dag}_{\text{L},\bq \ua}
\gamma_{\text{L},\ua}
e^{-i\phi/2}
)P_{n_{0}},
\\
&P_{n_{0}}
(
\lambda_{\text{R}}
\gamma_{\text{R},\da}
c_{\text{R},-\bk \da}
e^{i\phi/2}
)
(
\lambda_{\text{L}}
c^{\dag}_{\text{L},-\bq \da}
\gamma_{\text{L},\da}
e^{-i\phi/2}
)
(
\lambda_{\text{R}}
\gamma_{\text{R},\ua}
c_{\text{R},\bk \ua}
e^{i\phi/2}
)
(
\lambda_{\text{L}}
c^{\dag}_{\text{L},\bq \ua}
\gamma_{\text{L},\ua}
e^{-i\phi/2}
)P_{n_{0}},
\end{split}
\end{equation}
all evaluate to
\begin{equation}P_{n_0}
(
\gamma_{\text{R},\ua}
\gamma_{\text{L},\ua}
\gamma_{\text{R},\da}
\gamma_{\text{L},\da}
)
e^{i(\varphi_{\text{L}}-\varphi_{\text{R}})}
\lambda_{\text{R}}^{2}
\lambda_{\text{L}}^{2}
v_{\bq}
u_{\bk}
u_{\bq}
v_{\bk}
P_{n_0}.
\end{equation}
Moreover, they also share the same energy denominators. For that reason, once we combine them with the corresponding hermitian-conjugated sequences, multiply by the energy denominators and sum over all momenta, we arrive at the contribution
\begin{equation}
H'_{\text{eff},0}\equiv
-16P_{n_0}(
\gamma_{\text{R},\ua}
\gamma_{\text{L},\ua}
\gamma_{\text{R},\da}
\gamma_{\text{L},\da}
)
\lambda_{\text{R}}^{2}
\lambda_{\text{L}}^{2}
\cos(\varphi_{\text{L}}-\varphi_{\text{R}})
\sum_{\bk,\bq}
\frac{
v_{\bq}
u_{\bk}
u_{\bq}
v_{\bk}
}
{
(E_{\bq}+U)(E_{\bk}+E_{\bq})(E_{\bk}+U)
} P_{n_0}
\end{equation}
to the effective Hamiltonian. We can now rewrite the summation over the momenta in terms of an integral over the density of states. Then the contribution to the effective Hamiltonian given above simplifies to
\begin{equation}
\begin{split}
H'_{\text{eff},0}=
-P_{n_0}(
\gamma_{\text{R},\ua}
\gamma_{\text{L},\ua}
\gamma_{\text{R},\da}
\gamma_{\text{L},\da}
)J'_{0}\cos(\varphi_{\text{L}}-\varphi_{\text{R}})P_{n_0},
\end{split}
\end{equation}
where $J'_{0}$ was defined in Eq.~(11) of the main text.

As a second step, we consider the sequences of intermediate states given by
\begin{equation}
\begin{split}
&
P_{n_0}
(
\lambda_{\text{L}}
\gamma_{\text{L},\ua}
c_{\text{L},\bk \ua}
e^{i\phi/2}
)
(
\lambda_{\text{R}}
c^{\dag}_{\text{R},\bq \ua}
\gamma_{\text{R},\ua}
e^{-i\phi/2}
)
(
\lambda_{\text{R}}
c^{\dag}_{\text{R},-\bq \da}
\gamma_{\text{R},\da}
e^{-i\phi/2}
)
(
\lambda_{\text{L}}
\gamma_{\text{L},\da}
c_{\text{L},-\bk \da}
e^{i\phi/2}
)
P_{n_0},
\\
&
P_{n_0}
(
\lambda_{\text{L}}
\gamma_{\text{L},\ua}
c_{\text{L},\bk \ua}
e^{i\phi/2}
)
(
\lambda_{\text{R}}
c^{\dag}_{\text{R},-\bq \da}
\gamma_{\text{R},\da}
e^{-i\phi/2}
)
(
\lambda_{\text{R}}
c^{\dag}_{\text{R},\bq \ua}
\gamma_{\text{R},\ua}
e^{-i\phi/2}
)
(
\lambda_{\text{L}}
\gamma_{\text{L},\da}
c_{\text{L},-\bk \da}
e^{i\phi/2}
)
P_{n_0},
\\
&
P_{n_0}
(
\lambda_{\text{L}}
\gamma_{\text{L},\da}
c_{\text{L},-\bk \da}
e^{i\phi/2}
)
(
\lambda_{\text{R}}
c^{\dag}_{\text{R},\bq \ua}
\gamma_{\text{R},\ua}
e^{-i\phi/2}
)
(
\lambda_{\text{R}}
c^{\dag}_{\text{R},-\bq \da}
\gamma_{\text{R},\da}
e^{-i\phi/2}
)
(
\lambda_{\text{L}}
\gamma_{\text{L},\ua}
c_{\text{L},\bk \ua}
e^{i\phi/2}
)
P_{n_0},
\\
&
P_{n_0}
(
\lambda_{\text{L}}
\gamma_{\text{L},\da}
c_{\text{L},-\bk \da}
e^{i\phi/2}
)
(
\lambda_{\text{R}}
c^{\dag}_{\text{R},-\bq \da}
\gamma_{\text{R},\da}
e^{-i\phi/2}
)
(
\lambda_{\text{R}}
c^{\dag}_{\text{R},\bq \ua}
\gamma_{\text{R},\ua}
e^{-i\phi/2}
)
(
\lambda_{\text{L}}
\gamma_{\text{L},\ua}
c_{\text{L},\bk \ua}
e^{i\phi/2}
)
P_{n_0},
\end{split}
\end{equation}
and the corresponding sequences with $\text{L}\leftrightarrow\text{R}$. These sequences evaluate to
\begin{equation}P_{n_0}
(
\gamma_{\text{R},\ua}
\gamma_{\text{L},\ua}
\gamma_{\text{R},\da}
\gamma_{\text{L},\da}
)
e^{i(\varphi_{\text{L}}-\varphi_{\text{R}})}
\lambda_{\text{R}}^{2}
\lambda_{\text{L}}^{2}
v_{\bq}
u_{\bk}
u_{\bq}
v_{\bk}
P_{n_0},
\end{equation}
while the corresponding sequences with $\text{L}\leftrightarrow\text{R}$ evaluate to $P_{n_0}(
\gamma_{\text{R},\ua}
\gamma_{\text{L},\ua}
\gamma_{\text{R},\da}
\gamma_{\text{L},\da}
)
e^{-i(\varphi_{\text{L}}-\varphi_{\text{R}})}
\lambda_{\text{R}}^{2}
\lambda_{\text{L}}^{2}
v_{\bq}
u_{\bk}
u_{\bq}
v_{\bk}
P_{n_0}$. However, all of these sequences share the same energy denominators. Consequently,
after adding the hermitian-conjugated sequences, multiplying by the energy denominators and performing the summation
over all momenta, we find the following contribution to the effective Hamiltonian,
\begin{equation}
H'_{\text{eff},1}\equiv
-
16P_{n_0}(
\gamma_{\text{R},\ua}
\gamma_{\text{L},\ua}
\gamma_{\text{R},\da}
\gamma_{\text{L},\da}
)
\lambda_{\text{R}}^{2}
\lambda_{\text{L}}^{2}
\cos(\varphi_{\text{L}}-\varphi_{\text{R}})
\sum_{\bk,\bq}
\frac{
v_{\bq}
u_{\bk}
u_{\bq}
v_{\bk}
}
{
(E_{\bk}+U)^{2}(E_{\bk}+E_{\bq})
} P_{n_0}.
\end{equation}
By expressing the summation over the momenta in terms of an integral over the density of states, this contribution to the effective Hamiltonian can be rewritten as,
\begin{equation}
\begin{split}
H'_{\text{eff},1}=
-P_{n_0}(
\gamma_{\text{R},\ua}
\gamma_{\text{L},\ua}
\gamma_{\text{R},\da}
\gamma_{\text{L},\da}
)J'_{1}\cos(\varphi_{\text{L}}-\varphi_{\text{R}})P_{n_0},
\end{split}
\end{equation}
where $J'_{1}$ was defined in Eq.~(11) of the main text.

As a final step, we consider sequences of intermediate states in which we subsequently add and remove two units of charge on the TRI TSC. These sequences are given by,
\begin{equation}
\begin{split}
&
P_{n_0}
(
\lambda_{\text{R}}
c^{\dag}_{\text{R},\bq \ua}
\gamma_{\text{R},\ua}
e^{-i\phi/2}
)
(
\lambda_{\text{R}}
c^{\dag}_{\text{R},-\bq \da}
\gamma_{\text{R},\da}
e^{-i\phi/2}
)
(
\lambda_{\text{L}}
\gamma_{\text{L},\ua}
c_{\text{L},\bk \ua}
e^{i\phi/2}
)
(
\lambda_{\text{L}}
\gamma_{\text{L},\da}
c_{\text{L},-\bk \da}
e^{i\phi/2}
)
P_{n_0},
\\
&
P_{n_0}
(
\lambda_{\text{R}}
c^{\dag}_{\text{R},\bq \ua}
\gamma_{\text{R},\ua}
e^{-i\phi/2}
)
(
\lambda_{\text{R}}
c^{\dag}_{\text{R},-\bq \da}
\gamma_{\text{R},\da}
e^{-i\phi/2}
)
(
\lambda_{\text{L}}
\gamma_{\text{L},\da}
c_{\text{L},-\bk \da}
e^{i\phi/2}
)
(
\lambda_{\text{L}}
\gamma_{\text{L},\ua}
c_{\text{L},\bk \ua}
e^{i\phi/2}
)
P_{n_0},
\\
&
P_{n_0}
(
\lambda_{\text{R}}
c^{\dag}_{\text{R},-\bq \da}
\gamma_{\text{R},\da}
e^{-i\phi/2}
)
(
\lambda_{\text{R}}
c^{\dag}_{\text{R},\bq \ua}
\gamma_{\text{R},\ua}
e^{-i\phi/2}
)
(
\lambda_{\text{L}}
\gamma_{\text{L},\ua}
c_{\text{L},\bk \ua}
e^{i\phi/2}
)
(
\lambda_{\text{L}}
\gamma_{\text{L},\da}
c_{\text{L},-\bk \da}
e^{i\phi/2}
)
P_{n_0},
\\
&
P_{n_0}
(
\lambda_{\text{R}}
c^{\dag}_{\text{R},-\bq \da}
\gamma_{\text{R},\da}
e^{-i\phi/2}
)
(
\lambda_{\text{R}}
c^{\dag}_{\text{R},\bq \ua}
\gamma_{\text{R},\ua}
e^{-i\phi/2}
)
(
\lambda_{\text{L}}
\gamma_{\text{L},\da}
c_{\text{L},-\bk \da}
e^{i\phi/2}
)
(
\lambda_{\text{L}}
\gamma_{\text{L},\ua}
c_{\text{L},\bk \ua}
e^{i\phi/2}
)
P_{n_0},
\\
\\
&
P_{n_0}
(
\lambda_{\text{L}}
\gamma_{\text{L},\ua}
c_{\text{L},\bk \ua}
e^{i\phi/2}
)
(
\lambda_{\text{L}}
\gamma_{\text{L},\da}
c_{\text{L},-\bk \da}
e^{i\phi/2}
)
(
\lambda_{\text{R}}
c^{\dag}_{\text{R},\bq \ua}
\gamma_{\text{R},\ua}
e^{-i\phi/2}
)
(
\lambda_{\text{R}}
c^{\dag}_{\text{R},-\bq \da}
\gamma_{\text{R},\da}
e^{-i\phi/2}
)
P_{n_0},
\\
&
P_{n_0}
(
\lambda_{\text{L}}
\gamma_{\text{L},\ua}
c_{\text{L},\bk \ua}
e^{i\phi/2}
)
(
\lambda_{\text{L}}
\gamma_{\text{L},\da}
c_{\text{L},-\bk \da}
e^{i\phi/2}
)
(
\lambda_{\text{R}}
c^{\dag}_{\text{R},-\bq \da}
\gamma_{\text{R},\da}
e^{-i\phi/2}
)
(
\lambda_{\text{R}}
c^{\dag}_{\text{R},\bq \ua}
\gamma_{\text{R},\ua}
e^{-i\phi/2}
)
P_{n_0},
\\
&P_{n_0}
(
\lambda_{\text{L}}
\gamma_{\text{L},\da}
c_{\text{L},-\bk \da}
e^{i\phi/2}
)
(
\lambda_{\text{L}}
\gamma_{\text{L},\ua}
c_{\text{L},\bk \ua}
e^{i\phi/2}
)
(
\lambda_{\text{R}}
c^{\dag}_{\text{R},\bq \ua}
\gamma_{\text{R},\ua}
e^{-i\phi/2}
)
(
\lambda_{\text{R}}
c^{\dag}_{\text{R},-\bq \da}
\gamma_{\text{R},\da}
e^{-i\phi/2}
)
P_{n_0},
\\
&
P_{n_0}
(
\lambda_{\text{L}}
\gamma_{\text{L},\da}
c_{\text{L},-\bk \da}
e^{i\phi/2}
)
(
\lambda_{\text{L}}
\gamma_{\text{L},\ua}
c_{\text{L},\bk \ua}
e^{i\phi/2}
)
(
\lambda_{\text{R}}
c^{\dag}_{\text{R},-\bq \da}
\gamma_{\text{R},\da}
e^{-i\phi/2}
)
(
\lambda_{\text{R}}
c^{\dag}_{\text{R},\bq \ua}
\gamma_{\text{R},\ua}
e^{-i\phi/2}
)
P_{n_0}.
\end{split}
\end{equation}
All of these sequences evaluate to
\begin{equation}P_{n_0}
(
\gamma_{\text{R},\ua}
\gamma_{\text{L},\ua}
\gamma_{\text{R},\da}
\gamma_{\text{L},\da}
)
e^{i(\varphi_{\text{L}}-\varphi_{\text{R}})}
\lambda_{\text{R}}^{2}
\lambda_{\text{L}}^{2}
v_{\bq}
u_{\bk}
u_{\bq}
v_{\bk}
P_{n_0}.
\end{equation}
Moreover, they also all share the same energy denominators. Hence,
after adding the hermitian-conjugated sequences, multiplying by the energy denominators and performing the summation
over all momenta, we find that they lead to the following contribution to the effective Hamiltonian,
\begin{equation}
H'_{\text{eff},2}\equiv-
16P_{n_0}(
\gamma_{\text{R},\ua}
\gamma_{\text{L},\ua}
\gamma_{\text{R},\da}
\gamma_{\text{L},\da}
)
\lambda_{\text{R}}^{2}
\lambda_{\text{L}}^{2}
\cos(\varphi_{\text{L}}-\varphi_{\text{R}})
\sum_{\bk,\bq}
\frac{
v_{\bq}
u_{\bk}
u_{\bq}
v_{\bk}
}
{
(E_{\bq}+U)(4U)(E_{\bk}+U)
} P_{n_0}.
\end{equation}
By expressing the summation over the momenta in terms of an integral over the density of states, we note this contribution to the effective Hamiltonian can be written in the simplified form,
\begin{equation}
\begin{split}
H'_{\text{eff},2}=-
P_{n_0}
(
\gamma_{\text{R},\ua}
\gamma_{\text{L},\ua}
\gamma_{\text{R},\da}
\gamma_{\text{L},\da}
)J'_{2}\cos(\varphi_{\text{L}}-\varphi_{\text{R}})P_{n_0},
\end{split}
\end{equation}
where $J'_{2}$ was defined in Eq.~(11) of the main text. The full effective Hamiltonian is then given by
\begin{equation}
H'_{\text{eff}}=H'_{\text{eff},0}+H'_{\text{eff},1}+H'_{\text{eff},2}.
\end{equation}
For the case when the joint parity of two MKPs $\gamma_{\text{L},s}$, $\gamma_{\text{R},s}$ is equal
to the total island charge mod 2,
\begin{equation}
\gamma_{\text{R},\ua}
\gamma_{\text{L},\ua}
\gamma_{\text{R},\da}
\gamma_{\text{L},\da}
=
(-1)^{n_0},
\end{equation}
this result simplifies to
\begin{equation}
\begin{split}
H'_{\text{eff}}=
(-1)^{n_{0}+1}(J'_{0}+J'_{1}+J'_{2})\cos(\varphi_{\text{L}}-\varphi_{\text{R}})
=
(-1)^{n_{0}+1}J'\cos(\varphi_{\text{L}}-\varphi_{\text{R}})
,
\end{split}
\end{equation}
which concludes the derivation.

\end{widetext}

\end{document}